\documentclass[preprint,12pt,nofootinbib]{revtex4}
\usepackage{graphicx}
\usepackage{cancel}
\usepackage{textcomp}
\usepackage{amsmath, amssymb, amsfonts, latexsym, epsfig}
\usepackage{mathrsfs}
\usepackage{bm}
\usepackage{times}
\usepackage{epsfig}
\usepackage{color}
\usepackage{slashed}

\def\beq{\begin{equation}}
\def\eeq{\end{equation}}
\def\bea{\begin{eqnarray}}
\def\eea{\end{eqnarray}}

\begin{document}

\title{Pure leptonic proposal to $W^{+}W^{-}$ excess and neutrino mass generation}
\author{Hui Luo$^{1,2,3}$}
\author{Ming-xing Luo$^{1}$}
\author{Kai Wang$^{1}$}
\author{Tao Xu$^{1}$} 
\author{Guohuai Zhu$^{1}$}
\affiliation{$^{1}$~Zhejiang Institute of Modern Physics, Department of Physics, Zhejiang University, Hangzhou, Zhejiang 310027, China\\
$^{2}$~Dipartimento di Fisica ed Astronomia, Universit`a di Padova, Via Marzolo 8, 35131 Padova, Italy\\
$^{3}$~INFN, Sezione di Padova, Via Marzolo 8, 35131 Padova, Italy
}

\begin{abstract}
We investigate the TeV models for neutrino mass generation as candidate models to explain the recent 2$\sigma$ excess of leptonic $W^{+}W^{-}$ pair production at LHC.
Several models with singly charged exotic states that may explain the excess require light masses completely excluded by LEP experiments.  One possible model \cite{hitoshi} with new lepton doublets can fit the observation and evade all direct search bounds but with tuned Yukawa structure to satisfy lepton universality. The new exotic leptons $L^{\pm}$ decay into $L^{\pm}\to \ell^{\pm} \phi$ where $\phi$ is a light singlet scalar of $\cal O$(MeV) that decays into neutrinos. Drell-Yan production of $L^{+}L^{-}\to \ell^{+}\ell^{-}+\cancel{E}_{T}$ fits the excess and $L^{\pm}L^{0}\to \ell^{\pm}+\cancel{E}_{T}$ is completely buried in SM background. 

\end{abstract}
\maketitle
Very recently, both ATLAS and CMS found 2$\sigma$ excess in $W^{+}W^{-}$ pair measurements \cite{exp} while the $ZZ$ measurements
are more consistent with the SM predictions. The latest 
analysis on pure leptonic $W^{+}W^{-}$ based on 8 TeV LHC data from ATLAS and CMS are listed in Eq. \ref{number} as 
\bea
\sigma^{ATLAS@8TeV}_{W^{+}W^{-}}&=&71.4\pm 1.2(\text{stat.})\pm 4.5(\text{syst.})\pm 2.1 (\text{lumi.})\text{pb}\nonumber\\
\sigma^{CMS@8TeV}_{W^{+}W^{-}}&=& 69.9\pm 2.8 (\text{stat.}) \pm 5.6 (\text{syst.}) \pm 3.1 (\text{lumi.}) \text{pb}~.
\label{number}
\eea
The SM prediction \cite{ww} is 
\beq
\sigma^{SM}_{W^{+}W^{-}}=58.7 ^{+1.0}_{-1.1}(\text{PDF})^{+3.1}_{-2.7}(\text{total}) \text{pb}~.
\eeq
At Tevatron experiments,
both CDF and D$\cancel{0}$ also found central values significantly larger than the SM predictions but the error bars were
also large \cite{tevatron}. On the other hand, combined analysis of LEP II experiments \cite{lep} put stringent bounds on pure leptonic $W$ pair below $\sqrt{s}\le 206~\text{GeV}$ with  $R_{W^{+}W^{-}} = 0.995\pm 0.008$
which is the ratio of measured production cross section for $W^{+}W^{-}$ pair and the SM prediction.
There are attempts to explain the excess through new resummation calculation \cite{qcd}
but the excess has also generated several proposals based on supersymmetric models, in particular, light top squark in ``natural SUSY'' scenario \cite{susy}. 
The colored scalar production rate at 8 TeV LHC is of $\cal O$(10~pb).
With similar leptonic decay branching as Br($W^{-}\to \ell^{-}\bar{\nu}$), signal identical visible final states 
can well fake the leptonic $W$. Degeneracy condition in spectrum as $M_{\tilde{t}_{1}}-M_{\tilde{\chi}^{\pm}_{1}}\sim {\cal O}(\text{GeV})$ is also imposed to avoid the 
visible $b$-jet. On the other hand, the excess of leptonic $W^{+}W^{-}$ may also imply
the extension in leptonic sector,
in particular, TeV ``see-saw'' scenarios (also known as ``inverse see-saw'' sometimes) for neutrino mass generation
and, in this paper, we investigate the possibility of this pure leptonic approach.

Discovery of a 125~GeV Standard Model (SM)-like Higgs boson at the CERN Large Hadron Collider has dramatically improved our knowledge on mass generation for elementary particles in SM \cite{today}. However, clear evidence for physics beyond SM lies in experimental confirmation of sub-eV neutrino masses  based on distance/energy dependence measurements in various neutrino oscillation experiments \cite{nex}. Being complete neutral under unbroken gauge symmetry $SU(3)_{C}\times U(1)_{EM}$, neutrino can be Majorana fermion. Moreover, Majorana nature of neutrino also ensures the uniqueness of hyper-charge assignment predicted by gauge anomaly free conditions \cite{yanagidabook}. The total mass of neutrino states and upper bound on neutrino charge \footnote{Based on charge conservation assumption, the neutrino charge bound is further constrained as less than $10^{-21}e$~\cite{milli}. } are given in \cite{pdg} .
\beq
m_{\rm total}=\sum m_{{\nu}_{i}} \lesssim 0.24~\text{eV}, q_{\nu} \lesssim 10^{-15} e~.
\eeq
The most elegant proposal of neutrino mass generation  is the ``see-saw'' mechanism \cite{seesaw,so10} where the tiny but non-zero neutrino mass arises as a consequence of ultra-high scale (${\cal O}(\Lambda_{GUT}$)) physics and the mechanism can be naturally embedded into grand unification framework \cite{so10}. In addition, ``see-saw''
mechanism can naturally account for the observed baryon asymmetry of the universe from WMAP seven year results \cite{wmap} through ``leptogenesis''~\cite{leptogenesis}
\beq
Y_{B}\equiv \frac{\rho_{B}}{s} =(8.82\pm 0.23)\times 10^{-11},
\eeq
 where $\rho_{B}$ is the baryon number density and $s$ is 
 the entropy density of the universe. 

On the other hand, the ``see-saw'' mechanism is unlikely to be direct tested experimentally in near future. 
Heavy singlet fermion with strong Yukawa coupling to the Higgs boson leads to huge correction to the Higgs boson mass as $\delta m^{2}_{h} \simeq {m_{\nu}M^{3}_{R}}/{(2\pi v)^{2}}\log(q/M_{R})$~\cite{vissani}. Supersymmetry is then inevitable to stabilize the Higgs boson mass while low energy supersymmetry suffers severe direct search bounds at LHC. 
 Thermal leptogenesis also requires lower ``see-saw'' scale of ${\cal O}(10^{9}~\text{GeV})$ with smaller Dirac neutrino Yukawa couplings and large hierarchies in the right-handed neutrino masses~\cite{Buchmuller:2000as}.
Therefore, there are alternative proposals to generate neutrino masses within TeV.
Taking an effective field theory approach, neutrino mass in these models can be categorized into higher dimensional operator $(\phi^{n}/\Lambda^{n+1})\ell\ell h h$ with $n\neq 0$. For cut-off $\Lambda$ within TeV, $\phi$ is typically KeV-MeV known as inverse ``see-saw'' \cite{inverse,hitoshi}. Since the models mostly involve exotic physics only in leptonic sector, many of them have very distinguished predictions at hadron colliders with controlled background. 
 
 If the new exotic states only decay into leptonic final states as we discussed, one will only need 
${\cal O}(10\times (1/3)^{2}~\text{pb})$ to fake leptonic $W^{+}W^{-}$ final states. However, pure leptonic decaying scalar is typically excluded up to the LEP $\sqrt{s} \le 206$~GeV\footnote{Unless the spectrum is degenerate with extremely soft lepton final states.} while Drell-Yan production rate of scalar pair with mass greater than 103~GeV is much smaller than 1~pb and cannot account for the excess. 
Therefore, we focus on inverse ``see-saw'' (also known as TeV ``see-saw'' ) scenarios with Fermionic  extension
\footnote{A supersymmetric version of Zee model, Babu Model or Ma-model or Triplet-Higgs model~\cite{zongguo} also contain fermions with pure leptonic decay into leptons plus missing transverse energy $\ell+\cancel{E}_{T}$ which is singly-charged Higgsino in these models.}. 

Original inverse ``see-saw'' model only involves SM singlet fermions
$N$ which decay into di-lepton plus $\cancel{E}_{T}$ as $N\to \ell^{+}\ell^{-} \nu$. $N$ is produced at LHC as $pp\to \ell N$ which contribute to triple-lepton final states. In addition, the production is through light neutrino mixing which is also tiny. 

In TeV Type-III ``see-saw'' \cite{hexg}, a singly charged fermion from the
$SU(2)_{L}$ triplet $\Sigma^{\pm}$ can decay into $\ell+\cancel{E}$ as
\beq
\Sigma^{+}\to \nu W^{+}~~\text{with}~~W^{+}\to \ell^{+}\nu ,\Sigma^{+}\to \ell^{+} Z ~~\text{with}~~Z\to \nu\bar{\nu}~.
\eeq
However, first of all, $\Sigma^{+}$ does not always contribute to leptonic final states with only $80\% \times 3/9 + 20\%\times 20\%\simeq 30\%$ to $\ell^{+}+\cancel{E}_{T}$ final states which is similar
to $W$ decay. It then requires much larger production rate while Drell-Yan production at 8 TeV LHC for this non-colored fermion above the LEP bounds is not sufficient to account for the excess which has to be of $\cal O$(10~pb) in total. Secondly, 
$\Sigma^{\pm}\Sigma^{0}$ production which is larger than $\Sigma^{+}\Sigma^{-}$
pair production simultaneously predicts multi-lepton final states which suffers much severe experimental constraints. Therefore, a viable solution is to introduce new doublet fermion which is introduced in \cite{hitoshi}.

In \cite{hitoshi}, a pair of vector-like $SU(2)_{L}$ doublet fermions $L$ and $L^{c}$, a SM singlet $N$ are introduced.  
\beq
{\cal L}=YhNL+ML^{c} L+y\phi L^{c} l +{M_{N}\over 2}NN+h.c.
\eeq
where $l$ is the $SU(2)$ lepton doublet in SM, $h$ is the SM-like Higgs. 
\beq
L = \left(
\begin{array}{c}
  L^{0}\\
  L^{-} 
\end{array}
\right)_{L}
\eeq 
$\phi$ is a singlet scalar with mass of $\cal O$(MeV)
that decays into neutrino thus completely invisible 
in the detectors. 

Light neutrino mass arises as the dimensional-seven operator,
\beq
y^{2}Y^{2}\frac{\phi^{2}}{M_{N}M^{2}}ll h h~,
\label{opt}
\eeq
after integrating out the $L$ and $N$ fields. With $\langle \phi \rangle \sim {\cal O}$(KeV) and $M$, $M_{N}$ are all around ${\cal O}(10^{2}~\text{GeV}) $, one can easily obtain $m_{\nu}\sim {\cal O}$(eV).
Light singlet scalar $\phi$ participates in flavor physics processes and the $y$ should be carefully
chosen to be consistent with flavor physics constraints, for instance, $\mu\to e\gamma$ etc. For precision electroweak measurements, introduction of vector like doublets minimize the contribution to $S$-parameter but the
Yukawa couplings $Y$ to the SM-like Higgs should be less or equal to 
$0.2$ or so constrained by the $T$-parameter \cite{hitoshi}. On the other hand, these
couplings only appear in decays of exotic leptons and do not change the
qualitative feature of collider phenomenology. 

When $M_{N}+m_{h}> M$, $L\to N h$ decay is kinematically forbidden. 
In the $SU(2)$ limit, $L^{0}$ and $L^{-}$ are nearly degenerate and $L^{-}\to L^{0}\pi^{-}$ decay partial width is extremely small. However, as long as $y$ is
not highly suppressed,  
\beq
L^{\pm}\to \ell^{\pm} \phi~, L^{0}\to \nu \phi
\eeq
decay will dominate. The new neutral fermion $L^{0}$
is completely invisible. The singly charged exotic lepton decay into SM charged lepton plus $\cancel{E}_{T}$ which is identical to leptonic $W$ decay experimentally.  

The exotic fermions pair of $L^{\pm}$ and $L^{0}$ can be produced at LHC through gauge interaction
\beq
pp \to L^{+} L^{-} \to \ell^{+}\ell^{-} +\cancel{E}_{T}, ~~~~pp \to L^{\pm}L^{0} \to \ell^{\pm}+\cancel{E}_{T}~,
\eeq
where the di-lepton mode can be mis-identified as
leptonic $W^{+}W^{-}$ while the single-lepton mode
is also subject to test at direct search for $W^{\prime}$.  

Figure 1 shows the production rate
for $L^{+}L^{-}$ as well as the $L^{\pm}L^{0}$ at 8 TeV LHC.  
\begin{figure}[ht]
\includegraphics[width=8cm]{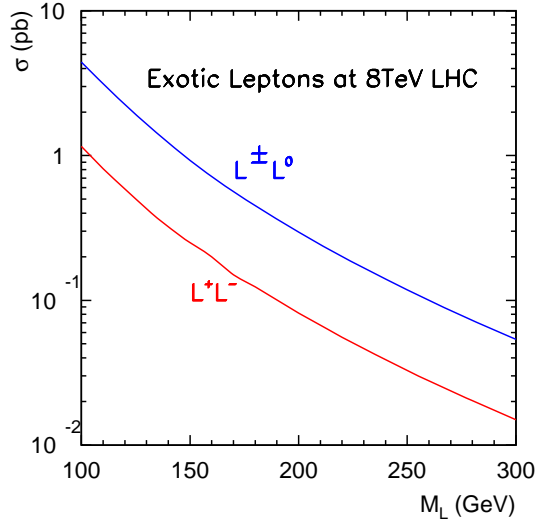}
\caption{Drell-Yan production rate for $L^{+}L^{-}$ pair $\sigma(pp\to L^{+}L^{-})$ at 8 TeV LHC. }
\label{Fig}
\end{figure}
As argued previously, to explain $\cal O$(10~pb) excess for $W^{+}W^{-}$, one needs pure leptonic final states to be of $\cal O$(pb).

Lepton universality is well tested at $W^{+}W^{-}$ pair measurements and
the excess has been observed in all lepton final states $e^{+}e^{-}$, $\mu^{+}\mu^{-}$ as well as $e^{\pm} \mu^{\mp}$. Therefore, it also put stringent constraints
on $L^{\pm}$ decay.  There are in principle three generations of $L^{\pm}$ and
their decays are determined by the $y_{ij}$. The exotic lepton $L^{\pm}$ decay into electron or muon 
\beq
L^{\pm}_{i} \to \ell^{\pm}_{j} \phi
\eeq
through
the Yukawa type of interactions $y_{ij} L^{c}_{i} l_{j} \phi$. 
The structure of Yukawa couplings $y_{ij}$ and the mass spectrum of $L_{i}$ 
may in principle affect the neutrino mass spectrum as in Eq.\ref{opt}.
However, this model contains much more freedoms than original ``see-saw''
mechanism \cite{seesaw} and therefore, $y_{ij}$ and $L$-mass matrix $M$
are less constrained.  
To keep the lepton universality, 
the simplest approach is that the lightest $L$ states dominated decay into $\tau^{\pm}\phi$
and the excess arises from $\tau^{\pm}\to \mu^{\pm}\nu\bar{\nu}$ or $\tau\to e^{\pm}\nu\bar{\nu}$ which makes about 17\% of $\tau$ decay each. The leptons from $\tau$ decay are typically softer than leptons directly from $W^{\pm}$ decay.
But, with larger mass, $\tau$-boost from $L$-decay is more significant than $\tau$s
from $W$-decay. In addition, leptons from left-handed polarized $\tau$ are also moving in the $\tau$-boosted direction. With all these factors taken into account, the lepton cut survival probability is expected to be higher than leptons from $\tau$ decaying from $W$s but less than the direct leptons from $W$ decay. Therefore, it would require much larger
production rate at the beginning. 

Therefore, even though with challenge, it is still possible to achieve the universality.  To illustrate the feature, in this paper, we discuss an oversimplified scenario with lepton universality for $L$-decay with decoupled $L_{2}$ and $L_{3}$. The Yukawa couplings $y_{ij}$ are taken to be 
\beq
y_{ij} \sim
\left(
\begin{array}{ccc}
 1 & 1  & 1  \\
 1 &  1 & 1  \\
 1 & 1  &  1
\end{array}
\right)~.
\eeq
The structure may suffer from constraints from lepton flavor violation tests and neutrino mass generation.
First of all, the Large mixing between different generation leptons may lead to large flavor violation mediated by $\phi$ and the model
may be severely constrained by bounds on $\mu\to e\gamma$
or $\mu\to e\gamma$ or so. But , with the contribution proportional to $y^{4}$, this bound can be evaded 
by making $y_{ij}$ smaller and this is irrelevant to collider phenomenology as long as the $L$ decay is not in meta-stable or long-lived range. Secondly, as we argued, Eq.\ref{opt}
connects $y_{ij}$ with neutrino mass matrix. However, 
even taken $y_{ij}$ as universal, there are as many degrees
of freedom as Type-I ``see-saw'' mechanism and one 
should be able to accommodate viable neutrino mass matrix just as in
Type-I ``see-saw'' mechanism.
If the Yukawa couplings $y_{ij}$ have universal structure. 
If $L_{2}$ and $L_{3}$ are of 250-300~GeV, the production
rate of $L_{2}$,$L_{3}$ pairs are only few percent of $L^{+}_{1}L^{-}_{1}$. $W^{\prime}$ search around this mass range is much less constrained due to background \cite{pdg}. With 
$L_{2}$, $L_{3}$ decoupled, we neglect the notation $i$ of $L_{i}$
in the following discussion and focus on the lightest $L_{i}$ production. 

We plot the normalized lepton $p_{T}$ distribution from $L^{\pm}\to \ell^{\pm}\phi$ decay of $L^{+}L^{-}$ pair in Fig. 2 in comparison with leptons in $W^{+}W^{-}$ production. In addition, lepton $p_{T}$ distribution in $L^{\pm}L^{0}$ production is very similar to its in $L^{+}L^{-}$. For comparison of
$W^{\prime}$ search, we also plot the lepton $p_{T}$ from single $W^{\prime}\to \ell^{\pm}\nu$ with $M_{W^{\prime}}=250$~GeV. 
\begin{figure}[ht]
\includegraphics[width=8cm]{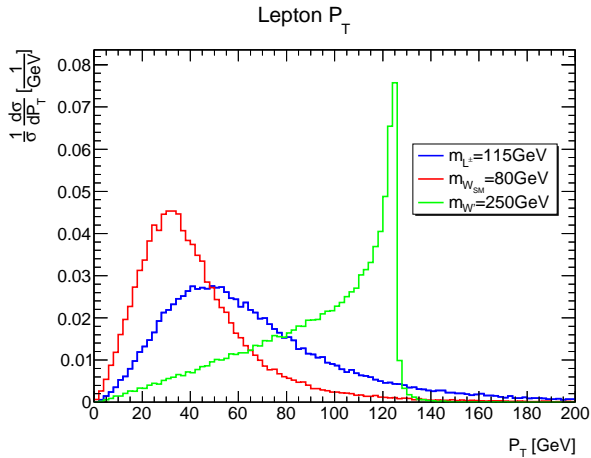}
\caption{$p_{T}$ distribution for leptons from SM $W$, $L$ of 115 GeV pair production and s-channel $W^{\prime}$ of 250~GeV.}
\label{Fig2}
\end{figure}
$L^{\pm}$ is slightly heavier than $W^{\pm}$ which results in harder lepton 
in its decay in comparison with $W$ decay. The $\cancel{E}_{T}$ in $L^{\pm}$
decay is also larger. 
Therefore, the lepton final states 
from $L^{+}L^{-}$ would have higher cuts survival probability. We compare
the cut survival probability in $L^{+}L^{-}$ with $W^{+}W^{-}$ and listed
them in Table I by implementing the ATLAS cuts \cite{exp}.
Final states are required to have exactly two leptons of opposite sign selected with 
the ATLAS defined criteria for isolated leptons. The leading lepton is required to 
have $p_{T} > 25$~GeV and the sub-leading lepton $p_{T} > 20$~GeV. 
To reduce the Drell-Yan di-lepton, $M_{\ell\ell}> 15$~GeV as well as 
$\mid M_{\ell\ell}-m_{Z}\mid> 15$~GeV. The study is performed by a modified version
of MadEvent \cite{madgraph}.
\begin{table}[ht]
\begin{tabular}{|c|c|c|c|c|c|c|}
\hline 
Cuts & $W^{+}W^{-}$  & 105~GeV & 110~GeV & 115~GeV & 120~GeV & 125~GeV\\
\hline 
\hline 
No $\cancel{p}_{T}$ cut & 0.170  & 0.180 & 0.174 & 0.170 & 0.167& 0.160\\
\hline 
$\cancel{p}_{T}>45$ & 0.055 &  0.120 & 0.121& 0.122 & 0.125 & 0.123\\
\hline 
$\cancel{p}_{T}>30$ & 0.096   & 0.147 & 0.144 & 0.143 & 0.145 & 0.140\\
\hline 
\end{tabular}
\caption{Cut survival probability $\epsilon$ for leptons decaying from SM $W^{\pm}$ and $L^{\pm}$. }
\end{table}
We use the ratio between survival probabilities of two channels, $\epsilon_{W^{+}W^{-}}/\epsilon_{L^{+}L^{-}}$, to estimate the required production rate for $L^{+}L^{-}$.
In principle, $L^{\pm}_{i}\to \ell^{\pm}_{j}\phi$ decay strongly depends on Yukawa couplings $y_{ij}$ which play important role in determining neutrino mass spectrum.
One can study implications on $L^{\pm}$ decays for different neutrino scenario, inverted hierarchy or normal hierarchy, etc, by studying correlation between Yukawa couplings $y_{ij}$ and $Y_{lm}$ and neutrino masses. These couplings are also strongly constrained by lepton flavor violation at the same time. 
To only illustrate the $W^{+}W^{-}$ excess feature, we do not make any further assumption
on neutrino masses except the estimated mass scale. Naively, the leading order production rate of $\sigma_{L^{+}L^{-}}$ can be estimate from 
\beq
\sigma_{LO} \simeq \Delta \sigma_{W^{+}W^{-}}\times \text{Br}(W^{\pm}\to \ell^{\pm}\nu) \times \text{Br}(W^{\pm}\to \ell^{\pm}\nu) \times {\epsilon_{W^{+}W^{-}} \over \epsilon_{L^{+}L^{-}}} / K_{QCD}
\eeq
where $K_{QCD}$ is the perturbative QCD $K$-factor for this Drell-Yan processes
which is about 1.6 for 8 TeV LHC Drell-Yan production of  weakly interacting particles
of $\cal O$(100~GeV). By taking a central ${\epsilon_{W^{+}W^{-}}/ \epsilon_{L^{+}L^{-}}}\sim 0.5$, the $\sigma_{LO}\sim 0.5$~pb which corresponds to $L^{\pm}$ of 125~GeV. The efficiencies are only estimated at the parton level and subjected to change when including real detector simulations. 

$L^{\pm}L^{0}\to \ell^{\pm}+\cancel{E}_{T}$ mode encounters direct search of $W^{\prime}$ at the LHC as single lepton plus missing transverse energy.
However, single $W$ production with $W^{\pm}\to \ell^{\pm}\nu$ at 8 TeV LHC
is about 5~nb with error bar 100~pb while $L^{\pm}L^{0}$ is only of $\cal O$(pb)
production rate. Lepton $p_{T}$ distribution in Fig. 2 also shows significant difference
between $L^{\pm}$ decay from heavy $W^{\prime}$. The latter one has a Jaccobian peak of $M_{W^{\prime}}/2$. The leptons from $\cal O$(100~GeV) $L^{\pm}$ state 
are more like leptons from $W$ decay so $L^{\pm}L^{0}$ is completely buried in
tails of SM $W$ background \cite{pdg}. 

Only left-handed SM lepton participates in $L^{-}$ decay
$L^{-}\to \tau^{-}\phi$. On the other hand, the SM $W^{-}$ decay
$W^{-}\to \tau^{-}\bar{\nu}$, $\tau^{-}$ is also left-handed polarized.
Hence, $\tau$-polarization cannot be used to distinguish the two
channels. 

\section*{Conclusion}
We study the TeV ``see-saw'' scenarios for neutrino mass generation to explain the recent 2$\sigma$ excess of leptonic $W^{+}W^{-}$ pair production at LHC and find
a model \cite{hitoshi} with one singlet neutrino plus additional vector-like lepton doublets can fit the observation and evade all direct search bounds. But the lepton universality test put stringent constraints over the Yukawa structure. 

\section*{Acknowledgement}
ML is supported by the National Science Foundation of China (11135006) and National Basic Research Program of China (2010CB833000). KW is supported in part, by the Zhejiang University Fundamental Research Funds for the Central Universities (2011QNA3017) and the National Science Foundation of China (11245002,11275168). GZ is supported in
part, by the National Science Foundation of China (11075139,
11375151), Zhejiang University Fundamental Research Funds for the Central Universities (2014QNA3006) the Program for New Century Excellent
Talents in University (Grant No. NCET-12-0480).

\end{document}